# Predicting Trends in $V_{OC}$ Through Rapid, Multimodal Characterization of State-of-the-Art p-i-n Perovskite Devices

Amy E. Louks,[1,2] Brandon T. Motes,[3] Anthony T. Troupe,[3] Axel F. Palmstrom,[2] Joseph J. Berry,[2,4] Dane W. deQuilettes[3*]

[1]Department of Chemistry, Colorado School of Mines, Golden, CO 80401 USA
[2]National Renewable Energy Laboratory, Golden, CO 80401 USA
[3]Optigon Inc, Somerville, MA 02143 USA
[4]Department of Physics and Renewable and Sustainable Energy Institute, University of Colorado Boulder, Boulder, CO 80309 USA

*Corresponding author: ddequilettes@optigon.us

**Abstract:**

Perovskite photovoltaic technologies are approaching commercial deployment, yet single junction and tandem architectures both still have significant room to improve power conversion efficiency and stability. The ability to perform rapid screening of material quality after altering processing conditions is critical to accelerating the optimization and commercialization of perovskite-based technologies. Currently, researchers utilize a wide range of stand-alone metrology tools to isolate sources of power loss throughout a device stack, which can be slow and labor intensive. Here, we demonstrate the use of a multimodal metrology approach to rapidly determine the maximum achievable and predicted open circuit voltages of > 100 perovskite devices during fabrication. Acquisition of these different data are facilitated by combining them into a single integrated measurement platform. We show that these data and automated analysis can be used to rapidly understand and ultimately predict quantitative trends in open circuit voltages of state-of-the-art devices architectures. The data and automated analysis workflow presented provides a reliable approach to quickly identify absorber and charge transport layer combinations that can lead to improved open circuit voltages.

**Main Text:**

Metal halide perovskite solar cells (PSCs) with the stoichiometry $ABX_3$ are a promising solar energy technology due to their ease of fabrication and excellent optoelectrical properties.[1] Single-junction perovskite solar cells have achieved power conversion efficiencies (PCEs) of 27.0% with multiple perovskite enabled tandems approaching or surpassing single junction limits (i.e. perovskite/silicon tandem solar cells achieving over 34% PCE).[2] Although impressive, these values are still far from their thermodynamic performance limits of 31% and 46%, respectively, and improvements in stability are greatly needed.[3,4] Further advances of PSCs will require accelerated materials characterization and an in depth understanding of factors limiting stability and efficiency such as electronic charge carrier recombination, transport, and carrier selectivity.[4] Indeed, identifying these factors will likely be critical to their large-scale deployment.[5-7]

To tackle the challenge of acquiring a more in depth understanding of the perovskite absorber layer, several groups have outlined a set of best practices to probe chemical structure and composition, electronic structure, optoelectronic properties, and device-relevant properties.[8] These sets of measurements typically include X-ray diffraction (XRD), X-ray photoelectron spectroscopy (XPS), current-voltage measurements (JV), secondary ion mass spectrometry (SIMS), ultraviolet photoemission spectroscopy (UPS) and, importantly, require careful calibration to avoid measurement artifacts. While the above techniques provide beneficial information about the material properties of PSCs, performing each measurement is time-intensive, can require completed devices with contacts, or are destructive. In the ideal scenario, all device relevant information can be accessed rapidly via non-contact (i.e. non-destructive) approaches and at any stage of device fabrication. Generally, this has been a key challenge in the photovoltaic (PV) community and several research groups and companies have built metrology tools to try and gain a holistic perspective of the factors limiting device performance.[9-12]

Both theoretical and experimental studies have identified material parameters that correlate strongly with PV device efficiency, which include the absorber layer absorption coefficient ($\alpha$), the charge carrier mobility ($\mu$), and the charge carrier lifetime ($\tau$).[13] Recent work has built on this understanding and included other key parameters such as the density of states, doping density, and static dielectric constant.[14,15] A significant breakthrough would be if all of these parameters could be rapidly measured to predict the device efficiency. In this study, we take a major step in addressing this challenge and use a multimodal, non-destructive tool that that autonomously measures key PV metrics through time resolved photoluminescence (TRPL), spectrally-resolved photoluminescence (SRPL), and transmission spectroscopy.[16-18] Each of these distinct measurements can provide critical information related to PV operation. For example, TRPL is a useful tool for measuring the charge carrier lifetime and transport properties of PV materials.[19] SRPL can be used to determine the PL quantum efficiency and recombination rates as well as the band-edge shape.[20] Transmission spectroscopy can be used to determine the absorption properties and thickness of the material. Combining the three of these techniques into one measurement tool can provide a more holistic evaluation of PV materials used in a device stack.[21]

Figure 1 shows an overview of the feedback cycle demonstrated in this study. First, we explore a set of different processing conditions for device fabrication, which are separated into 5 unique batches. Next, we characterize semi-fabricate samples midway through device completion through automated characterization. We use these data as inputs into an automated analysis pipeline to predict metrics relevant to device performance, such as the predicted or implied open circuit voltage. Last, we reach a "go" or "no go" decision point for completing devices that is based on the predicted metrics. Ultimately, the goal is to increase the learning rate for developing new processing conditions and device architectures.[22]

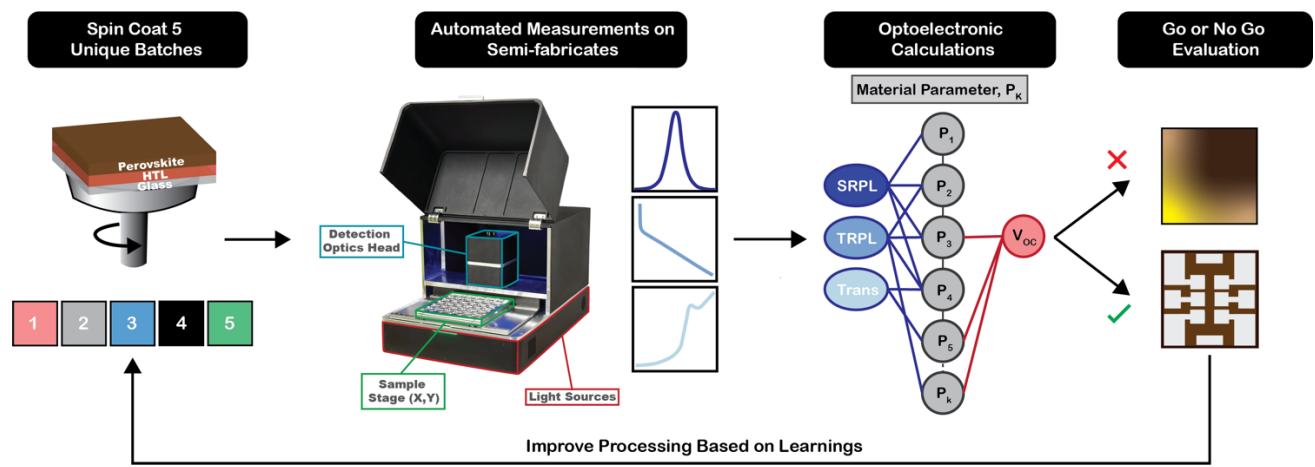

**Figure 1:** Flow diagram showing key aspects of this study which include spin coating of 5 unique sample batches (120 total devices); automated measurements using a multimodal tool with spectrally-resolved photoluminescence (SRPL), time-resolved PL (TRPL), and transmission (Trans) – 360 total optical

measurements; calculation of optoelectronic properties and device metrics, such as implied $V_{OC}$; and finally a decision point to complete a device or discard the samples.

A key goal of this study is to determine if a combination of rapid, non-contact measurements can be used to predict device performance metrics before devices are completed. Recently, Zhang *et al.* began validating this concept by using a neural net to predict electrical properties of completed devices by only using optical measurements as model inputs.[23] This study reported promising predictions especially for degraded devices, although physical interpretability of factors driving differences in performance were obscured due to the machine learning ("black box") approach. First-principles, physics-based predictions on semi-fabricates could provide additional insights such as the root-cause of underperformance and strategies to optimize material combinations. Therefore, we focus on developing a physics-informed approach that leverages the rapid data generation rate of the multimodal tool introduced in this work. Overall, predicting device performance before completion provides important time, cost, and material usage savings that could rapidly accelerate the development of PSCs.

Here, we test 5 separate methods for preparing devices and specifically focus on modifying two common aspects for device optimization: the perovskite active layer and hole transport layer. We use a control architecture of an inverted perovskite stack with the p-i-n architecture of indium-doped tin oxide (ITO)/Nickel Oxide (NiO$_X$)/FA$_{0.87}$MA$_{0.08}$Cs$_{0.05}$Pb(I$_{0.92}$Br$_{0.08}$)$_3$/Carbon-60 fullerene (C60)/bathocuproine (BCP)/Silver (Ag).[24] For hole transport layer modifications, we explore bilayers of NiO$_X$ with Poly[bis(4-phenyl)(2,4,6-trimethylphenyl)amine] (PTAA) and [2-(9*H*-carbazol-9-yl)ethyl]phosphonic Acid (2PACz) as they are known to lead to improvements in device $V_{OC}$.[25-28] For active layer modifications, we change the annealing time of the perovskite layer. Optimizing the perovskite annealing time is important to achieving desirable film microstructure, crystallinity, and background hole concentration.[29] Annealing the films too long or for too little time can result in lower PCEs and a noticeable change in the $V_{OC}$ of the devices due to increased charge recombination in the device.[30] We use this concept in our study and anneal the perovskite absorber layer for 15 minutes under or over the optimal anneal time compared to our standard recipe.

Figure 2 shows an overview of sample preparation, sample exchange, and device fabrication strategy deployed in this study. The National Renewable Energy Laboratory (NREL) team first fabricated a set of half-completed twin samples. NREL then kept one set and the other set was sent to Optigon. Optigon performed multimodal measurements at the locations where contacts would eventually be evaporated and then sent the samples back to NREL where they were made into full devices followed by measurement of current-voltage (JV) characteristics. In total, we prepared 4 sample replicates for each device batch (i.e. 20 samples = 4 samples x 5 sample batch variations). Each sample has 6 device pads, where we took 3 multimodal measurements for a total of 360 unique optical measurements (i.e. 360 measurements = 3 multimodal x 6 device pads x 20 samples). The control twin samples that were held at NREL were also completed into devices to test for possible degradation during transit. Figure S1 and S2 show the JV curves and $V_{OC}$ trends for each sample batch variation. Importantly, we observe similar trends in $V_{OC}$ for samples that stayed at NREL versus ones that were shipped to Optigon.

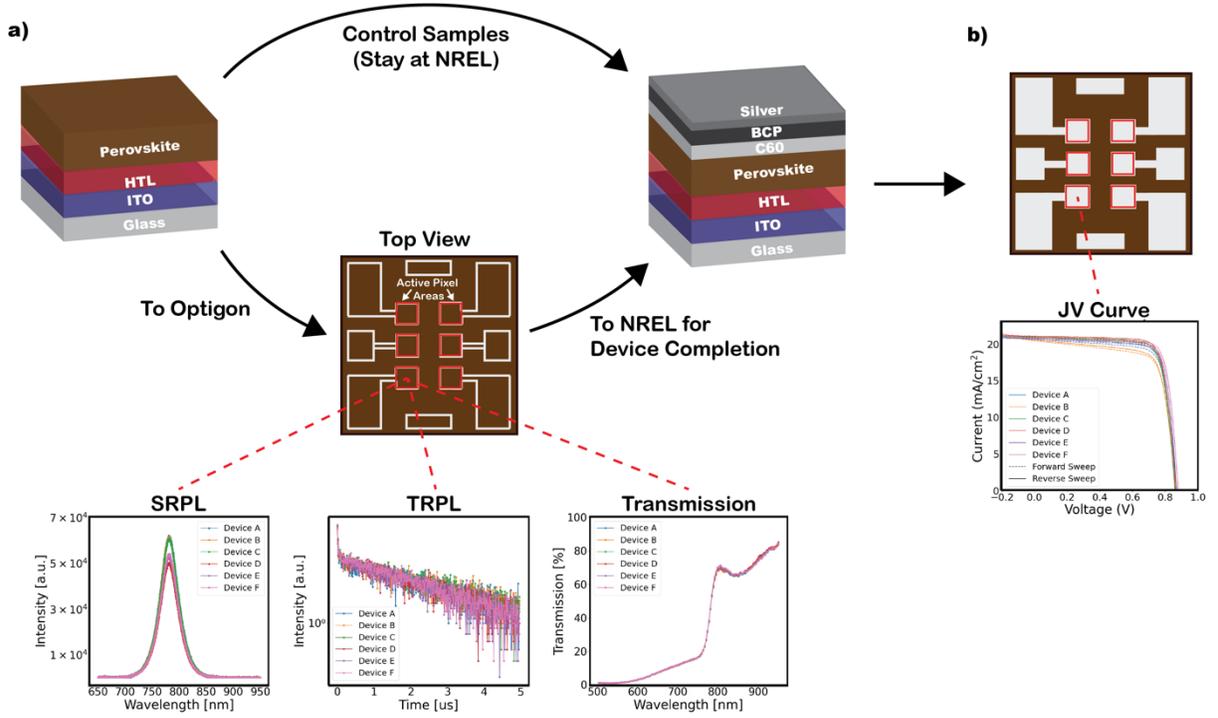

**Figure 2:** (a) Flow diagram of the sample transfer of semi-fabricate devices from NREL to Optigon. Optigon measured each sample by a rapid, multimodal optical tool in the area outlined with red, where the metal contact pads would eventually be deposited. The films were then sent back to NREL to be completed into full devices. Twin control films held at NREL were immediately made into devices and measured. The bottom of a) shows an example of a multimodal optical data set collected for one device on one sample. (b) An example set of JV curves after device completion.

To accurately predict the $V_{OC}$ of a device, it is important to know the real optical response of the device and not just the bandgap energy. Importantly, film thickness and bandtailing can have a major impact on the external radiative dark saturation current and upper limit of achievable voltage.[32] The $V_{OC}$ of a PV device in the radiative limit (i.e. thermodynamic limit with no non-radiative recombination) is determined by the optical response of the device at each wavelength and temperature.[33] One reliable method to determine the radiative $V_{OC}$ ($V_{OC,rad}$) is to measure the external quantum efficiency of the device, especially through the band tail, and then use detailed balance theory. The $V_{OC,rad}$ is determined by the short circuit current density ($J_{SC}$), external radiative dark saturation current ($J_{0rad,ext}$) and the thermal voltage as shown in Equation 1.

$$V_{OC}^{rad} = \frac{kT}{q}\ln\left[\frac{J_{SC}}{J_0^{rad,ext}}\right] = \frac{kT}{q}\ln\left[\frac{J_{SC}}{P_{esc}J_0^{rad,int}}\right] \quad \text{(Equation 1)}$$

Where $k$ is the Boltzmann constant, $T$ is the cell temperature, $J_{0rad,int}$ is the internal radiative saturation current, and $P_{esc}$ is the mean probability of photon escape from the film.[3]

Here, we determine the theoretical short circuit current density and dark saturation current density based on an approximated absorptivity spectrum. The absorptivity spectrum includes both the measured transmission data and the absorption band tail determined from the PL spectrum. This approach is similar to measuring the EQE and accounting for bandtailing using the electroluminescence (EL) spectrum.[33] Ideally, a PSC absorption spectrum has a steep absorptance onset and a sharp shoulder near the band gap energy of the film. Unfortunately, sub-bandgap absorption can be difficult to detect due to issues such as infrared scattering.[34] However, by using spectral PL and applying the reciprocity relationship we determine the bandtailing from the low energy shoulder of the PL spectra.[35, 36] We note that we do not account for reflectance or scattering from the sample, which are often required to get a fully accurate absorption coefficient. We report an Urbach energy of ~ 15 meV, which matches the expected values for this composition based on typical literature values.[35] Figure 3b show the absorption coefficient spectrum calculated by using the transmission and PL data for a single sample in this study.

In addition to the spectral PL and transmission, we measured the TRPL for every device pad on each sample. Figure 3c shows the low fluence TRPL decay trace for a single sample in this study. Because this is a measurement performed on a half-stack, we use the second decay component to represent the non-radiative recombination rate which includes bulk, surface, and interfacial recombination.[37] We use this non-radiative lifetime to simulate the PL quantum efficiency (PLQE) in the steady-state (Figure 3d) using a set of coupled differential equations consistent with previous work.[38, 39] Figure 3 shows an example data set and fitting of one device, we performed these measurements and automated modeling for every device pad on every sample (i.e. 120 total).[38, 39]

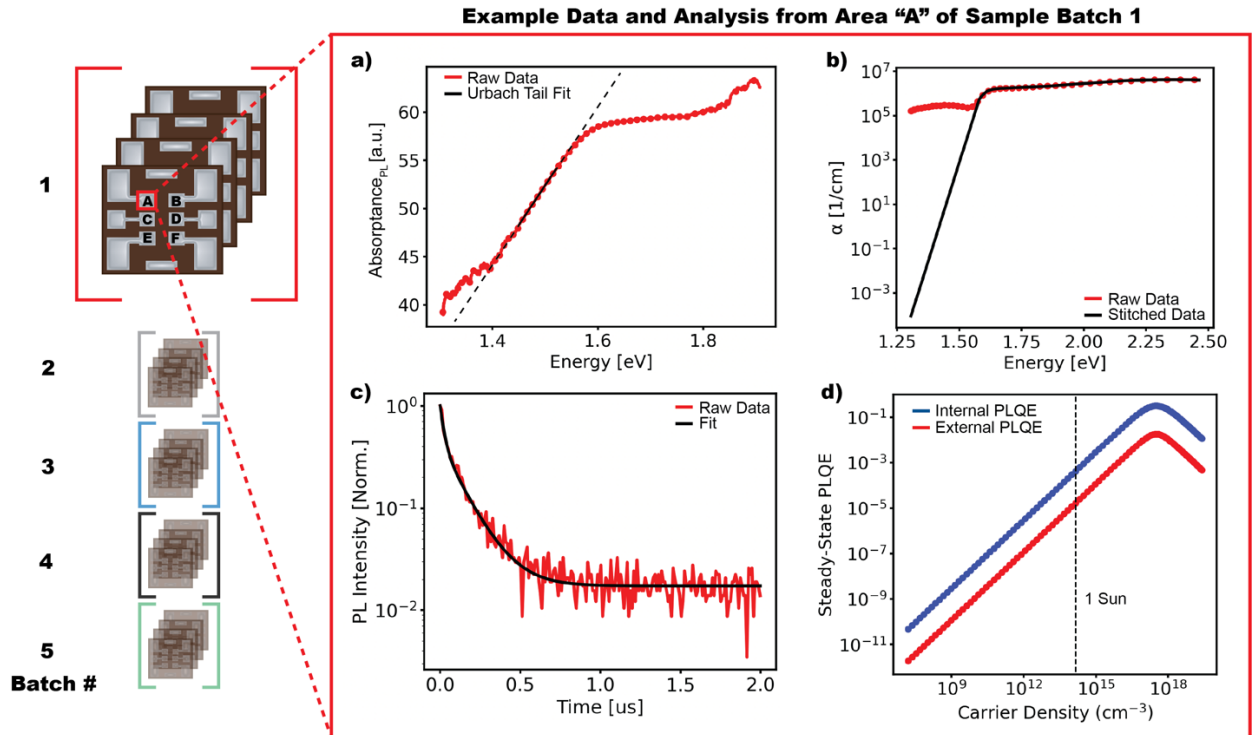

**Figure 3:** Example data set and analysis performed on a single device area of one sample batch. (a) Band-edge absorptivity function determined by applying the reciprocity relationship to the perovskite photoluminescence (PL) spectra. The solid black line shows the low energy Urbach tail fit over a specific energy range and the black dashed line is an extrapolation for easier visualization. (b) Absorption coefficient data determined from an absolute transmission measurement along with the stitched Urbach tail fit from (a). (c) Low fluence time-resolved PL decay trace of a perovskite semi-fabricate before completion of the device. The longer decay component is used to estimate the non-radiative recombination rate of the perovskite stack. (d) Simulated PL quantum efficiency (PLQE) versus carrier density using the non-radiative recombination rate from (c) as an input. The dotted line shows the expected PLQE at AM1.5 generation rate and the steady state carrier density. Each of these measurements were performed on 120 device pads across 5 sample batches.

Next, we calculate the predicted or implied $V_{OC}$ using the radiative $V_{OC}$ and the nonradiative $V_{OC}$ contribution as shown in equation 2.[40]

$$V_{OC} = V_{OC}^{rad} - kT|\ln(PLQE)| \quad \text{(Equation 2)}$$

Figure 4a shows the calculated PLQE for the 5 different sample variations explored in this study. We observe large differences in the calculated external PLQE for each sample set, especially for the different HTL combinations, indicative of variations in non-radiative loss. Figure 4b shows the radiative $V_{OC}$, which does not change significantly for each sample set, indicating that the bandgap and Urbach energy of the perovskite layer are not impacted by the different processing conditions. Figure 4b also shows the predicted $V_{OC}$ based on Equation 2. Notably, changing the HTL to a bilayer with PTAA or a SAM-like layer results in a ~150 mV increase in the predicted $V_{OC}$ relative to the control.

We observe only small changes in the predicted $V_{OC}$ for active layers annealed at different conditions. For the $NiO_x$/PTAA bilayer compared to the SAM-like layer, we predict a slightly lower predicted $V_{OC}$ for the SAM-like layer. This suggests that the most significant factor leading to differences in predicted $V_{OC}$ is the HTL/perovskite absorber layer interface rather than variations in the absorber layer. Figure 4b shows the $V_{OC}$ measured from JV curves on completed devices. Importantly, we observe similar trends across the sample sets when comparing to the predicted $V_{OC}$ calculations especially in separating the HTL modifications and small difference between the $NiO_x$/PTAA bilayer and SAM-like processing conditions.

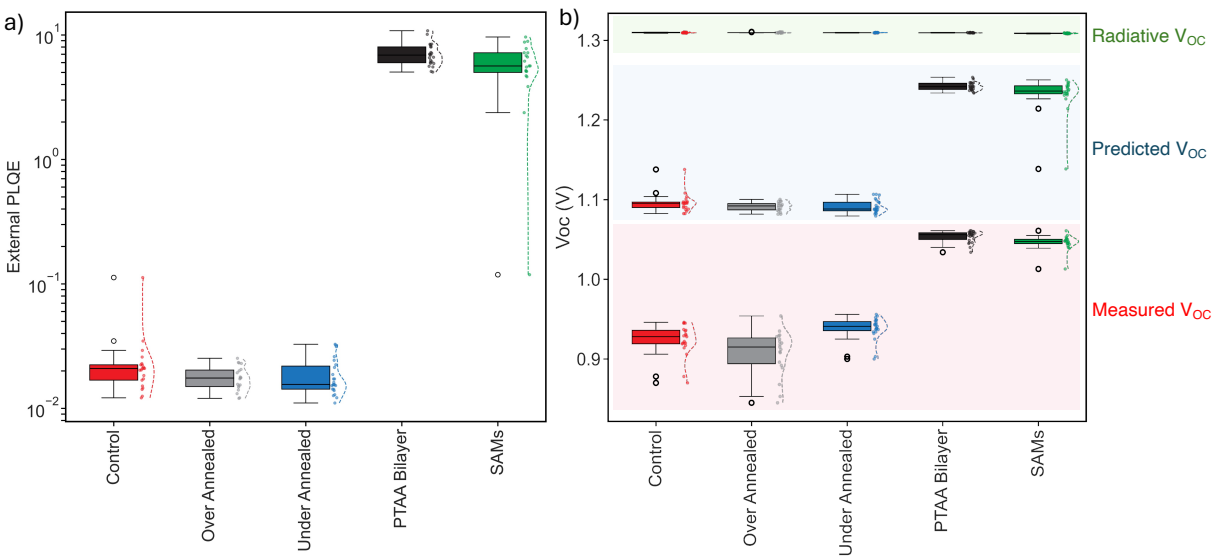

**Figure 4:** Box and whisker plot of the (a) calculated external PLQE's determined from the measured non-radiative recombination rates. (b) The calculated radiative open circuit voltages (maximum theoretical) for each type of sample. These values were calculated from the absorptivity spectrum determined by stitching

the PL determined Urbach tail to the measured transmission data. Also included is the predicted $V_{OC}$ based on the calculated PLQE values in (a) and the actual $V_{OC}$ values measured from current-voltage (JV) sweeps. For all box and whisker plots the sample size, n = 24, and the center line represents, median; box limits, upper and lower quartile; whiskers, 1.5x interquartile range; open black circles data points, outliers. Importantly, the predicted $V_{OC}$ trends from non-contact measurements performed on semi-fabricates correlate well with measured $V_{OC}$ values of completed devices.

A direct comparison of the implied $V_{OC}$ and actual $V_{OC}$ in Figure 4b shows that the measurements collected with the multimodal tool along with data analysis leads to good predictions in qualitative $V_{OC}$ trends for this sample set. The absolute offset between the predicted and measured $V_{OC}$ values are approximately 170 mV across all sets of devices (see Figure S3), suggesting that the same non-radiative loss is introduced by the deposition of the top charge transport layer and contacts (C60/BCP/Ag) upon completion of the device. This voltage loss is generally consistent with the 80 to 200 mV reported by others after deposition of C60.[41-44] We note that there may be other sources of loss, such as shunt pathways decreasing $V_{OC}$, but we can set an upper limit for additional non-radiative loss from the perovskite/ETL interface at ~170 mV.

Through this analysis, we can quantify the non-radiative voltage loss through various steps of device fabrication for each sample set. For example, in the control sample the $V_{OC,rad}$ is 1.31 V, which then decreases to a predicted $V_{OC}$ 1.09 V (determined from rapid optical measurements) for the semi-fabricate device. This suggests that ~220 mV of non-radiative loss originates from the perovskite and its interface with the HTL, $NiO_x$. The final device $V_{OC}$ for the control is 0.92 V (determined from JV measurements) after the deposition of C60/BCP/Ag, where an additional ~170 mV of non-radiative loss originates after the completion of the device, which most likely originates from the perovskite/ETL interface.

In conclusion, we use an automated, multimodal metrology tool to predict trends in $V_{OC}$ from semi-fabricate perovskite solar cell devices. The tool performs a set of 3 non-contact, optical measurements (spectrally-resolved PL, time-resolved PL, and transmission) in order to collect a holistic data set that captures both absorption and recombination processes. Automated measurements and analysis were conducted on 5 different sample sets exploring different perovskite processing and hole transport layer conditions. In total, 360 unique optical measurements were collected on half completed devices that were eventually completed into devices to perform correlations between optical and electrical measurements.

We report good predictions in the $V_{OC}$ trends in each perovskite device, indicating differences between each sample set in this study is driven by the HTL/perovskite interface. Through calculation of the maximum achievable $V_{OC}$ as well as optical and electrical measurements through the device fabrication process, we are able to isolate the origins of non-radiative voltage loss throughout the device stack. Notably, nonradiative recombination within the perovskite and at the perovskite/HTL interface leads to 60-200 mV loss in voltage and the perovskite/ETL interface leads to ≤ 170 mV loss in voltage. Importantly, we observed the lowest voltage losses in the devices that use a $NiO_x$/PTAA bilayer, showing that the bilayer can reduce the voltage loss by > 100 mV compared to semi-fabricates without PTAA.

Ultimately, this multi-modal measurement tool provides rapid feedback of the quality of samples during the device fabrication process. Predicting trends in $V_{OC}$ is a great first step for predicting PSC performance during fabrication and it is expected that the modeling can be expanded to predict the short-circuit current density ($J_{SC}$), fill factor (FF), and device PCE.

**Supporting Information Summary**

Experimental sections describing materials and device fabrication, electrical and optical measurements, and the sample exchange protocol between Optigon and NREL. Supporting Figures include plots of JV characteristics for different device batches.

**Author Contributions**

D.W.D., B.T.M., and A.T.T. designed and built the multimodal measurement tool. A.E.L., D.W.D., and J.J.B. conceived the experimental plan for utilizing the multimodal tool to study perovskite devices. A.E.L. made the perovskite samples and devices with help from A.F.P. D.W.D. and B.T.M performed all measurements on the semi-fabricated devices and wrote the multimodal data analysis code for $V_{OC}$ predictions. A.E.L. drafted the first version of the manuscript with input and edits from D.W.D. D.W.D., A.F.P., and J.J.B. supervised the research and all authors provided feedback on the manuscript.

**Notes**

D.W.D., B.T.M., and A.T.T. are co-founders of Optigon Inc., a US company developing metrology tools for the photovoltaics industry and other energy materials. The other authors declare no competing interests.

**Acknowledgements**

D.W.D., B.T.M., and A.T.T. acknowledge support from the U.S. Department of Energy's Office of Energy Efficiency and Renewable Energy (EERE) under the Solar Energy Technologies Office award number DE-EE0009838. A.E.L., A.F.P., and J.J.B. acknowledge support of the National Renewable Energy Laboratory, operated by Alliance for Sustainable Energy, LLC, for the U.S. Department of Energy (DOE) under Contract No. DE-AC36-08GO28308. This material is based upon work supported by the U.S. Department of Energy's Office of Energy Efficiency and Renewable Energy under the Solar Energy Technologies Office Award Number DE-052776. The views expressed in the article do not necessarily represent the views of the DOE or the U.S. Government. The U.S. Government retains, and the publisher, by accepting the article for publication, acknowledges that the U.S. Government retains a nonexclusive, paid-up, irrevocable, worldwide license to publish or reproduce the published form of this work, or allow others to do so, for U.S. Government purposes.

# Supporting Information: Predicting Trends in V$_{OC}$ Through Rapid, Multimodal Characterization of State-of-the-Art p-i-n Perovskite Devices

Amy E. Louks,[1,2] Brandon T. Motes,[3] Anthony T. Troupe,[3] Axel F. Palmstrom,[2] Joseph J. Berry,[2,4] Dane W. deQuilettes[3*]

[1]Department of Chemistry, Colorado School of Mines, Golden, CO 80401 USA
[2]National Renewable Energy Laboratory, Golden, CO 80401 USA
[3]Optigon Inc, Somerville, MA 02143 USA
[4]Department of Physics and Renewable and Sustainable Energy Institute, University of Colorado Boulder, Boulder, CO 80309 USA

*Corresponding author: ddequilettes@optigon.us

## Materials and Device Preparation

*Substrate and Chemical Information*

The substrates used were 25 x 25 mm patterned indium doped tin oxide coated soda lime glass (<20 Ω/sq) obtained from Colorado Concept Coatings (CCC). Poly[bis(4-phenyl)(2,4,6-trimethylphenyl)amine] (PTAA), toluene (anhydrous, 99.8%), cesium iodide (CsI, 99.999%), N,N-dimethylformamide (DMF, anhydrous, 99.8%), 1-methyl-2-pyrrolidone (NMP, anhydrous, 99.5%) were purchased from Sigma-Aldrich. Lead(II) iodide (PbI$_2$, 99.99%), [2-(3,6-dimethyl-9H-carbazol-9-yl)ethyl]phosphonic acid (Me-2PACz, >99.0%), and bathocuproine (BCP, purified by sublimation) were purchased from TCI Chemicals. Carbon 60 (C$_{60}$, <99.5%) was purchased from Luminescence Technology Corp. Methylammonium bromide (MABr) and formamidinium iodide (FAI) were purchased from Greatcell Solar Materials. The nickel oxide target (NiOx, 99.9%) and silver pellets (Ag, 99.9%) were purchased from Kurt J. Lesker.

*Sputtering*

ITO substrates were UV-ozoned and then moved into a Denton sputter coater to deposit nickel oxide via RF sputtering. Prior to the deposition step, the target was conditioned for 1000 s followed by a 400 s deposition process. The sputtering pressure was 25 mTorr with a 1:1 Ar/Ar:O$_2$ gas ratio at 60 W, resulting in a NiOx film with a thickness of 5 nm. The NiO$_x$ films were annealed for 10 min at 300 °C immediately before they were transferred into a glovebox used for active layer deposition.

*Hole Transport Layer Solution with PTAA*

A solution of poly[bis(4-phenyl)(2,4,6-trimethylphenyl)amine] (PTAA) was prepared by dissolving 2 mg of in 1 ml of toluene. The PTAA was deposited onto the NiOx layer with a spin coater at 6000 rpm for 30 seconds inside a nitrogen glovebox.

*Hole Transport Layer Solution with Self-Assembling Monolayer (SAM) Precursor*

A solution of Me-2PACz was prepared by dissolving 0.5 mg in 1 ml ethanol. The solution was spin coated onto ITO at 3000 rpm for 30 seconds in a nitrogen glovebox, then annealed at 100 °C for 10 minutes on a hot plate.

*Perovskite Absorber Layer*

The $FA_{0.87}MA_{0.08}Cs_{0.05}Pb(I_{0.92}Br_{0.08})_3$ precursor solution was prepared by dissolving a total of 1.3 M $Pb^{2+}$ calculated by finding the moles of $Pb^{2+}$ in a v/v = 12/88 NMP/DMF mixture, the salt weights are outlined in our previous work.[1] Once the desired HTL was deposited, the absorber layer was deposited in a two-step continuous nitrogen quench spin coating process. The first step is at a rate of 2000 rpm for 10 seconds followed by a rate of 6000 rpm for 24 seconds where a nitrogen blast is applied to the surface of the substrate for 15 seconds beginning 1 second into the second step. Once the spin coating process has completed, the film is then annealed for 30 minutes at 100 °C.

*Contact Layer Evaporation*

For device completion, the electron transport layer and contact layers were deposited with an Angstrom evaporator. On top of the active layer, 25 nm of C60 and then 6 nm of BCP were thermally deposited over the entire area. Finally, 100 nm of silver was thermally deposited through a metal mask.

*Sample Exchange/Transportation*

The samples exchanged with Optigon were finished through the absorber layer. For shipping, the samples were loaded into a 3D-printed plastic sample holder and put into a jar with a small amount of desiccant that was sealed in nitrogen. A small vacuum was pulled to ensure the jar was airtight. Next, the jar was vacuum sealed in a bag, bubble wrapped, and shipped. At Optigon, samples were stored in a desiccant cabinet before and after multimodal measurements, then immediately sent back to NREL for device completion.

**Electrical and Optical Measurements**

*JV Scans*

J-V scans were measured in an inert nitrogen glovebox with a Keithley 2450 source-measure unit. Sweeps were taken from -0.2 V to 1.2 V at a rate of 0.8 V/s. A class AAA LED G2V solar simulator calibrated with an AM1.5G filter was used for illumination. There was a small fan placed near the sample stage to keep air flow over the sample and to reduce heating effects of the lamp during measurements.

*Optigon Prism Tool*

The Prism tool is a rapid, automated, multimodal characterization tool commercialized by Optigon, Inc. Three measurements were deployed in this study which include:

*Transmission Measurements*

A broadband LED (420-950 nm) was used to illuminate the sample and signal was directed to a spectrophotometer both controlled with custom-built software. Similar to a standard UV-Vis measurement, a reference spectrum of the source emission through a glass slide without the perovskite was first measured. Next the perovskite sample was measured, and the transmission was calculated from the ratio of the two spectra (i.e. %T = ($I_{sample}$/$I_{reference}$) *100%).

*Spectrally-Resolved Photoluminescence*

A 405 nm continuous-wave diode laser was used to photoexcite the samples. The sample PL emission was filtered through a set of 500 nm long pass filters and directed to a spectrophotometer and data was collected using custom-built software.

*Time-Resolved Photoluminescence*

A 405 nm pulsed diode laser was used to photoexcite samples with a repetition rate of 200 kHz. The sample PL emission was filtered through a set of 700 nm long pass filters and directed to a single photon avalanche photodiode. Photon arrival times were time-tagged using a time-correlated single photon counter and data was collected using custom-built software.

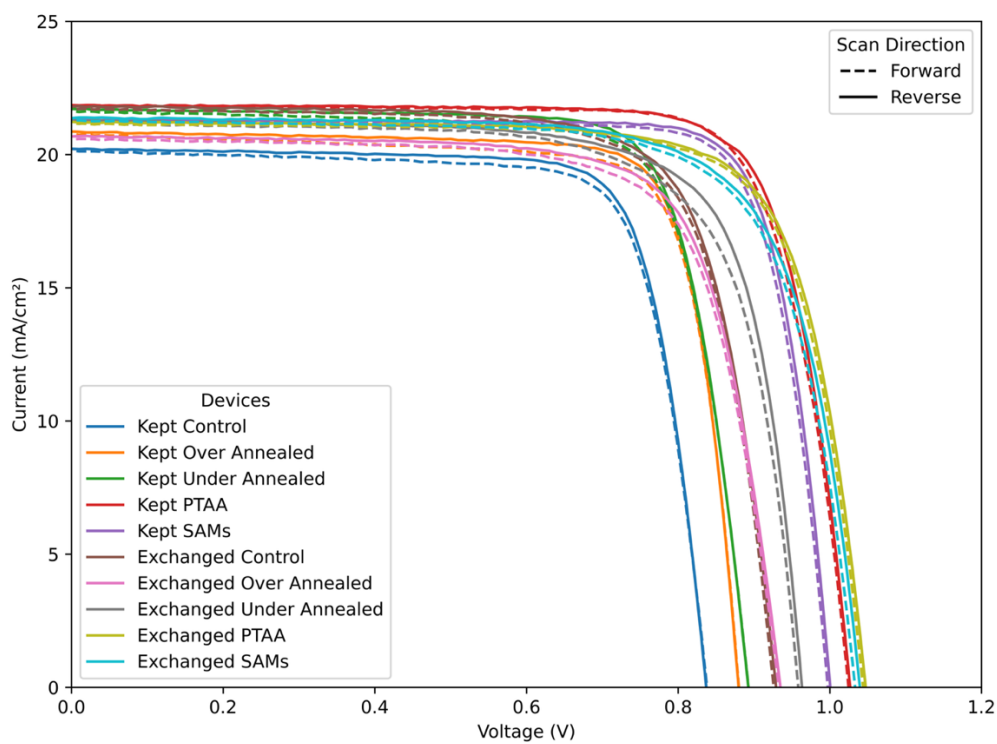

**Figure S1.** Champion JV scans for each material processing condition, including twin control samples kept at NREL compared to devices exchanged with Optigon. Forward scans are represented with a dashed line and reverse scans are represented with a solid line.

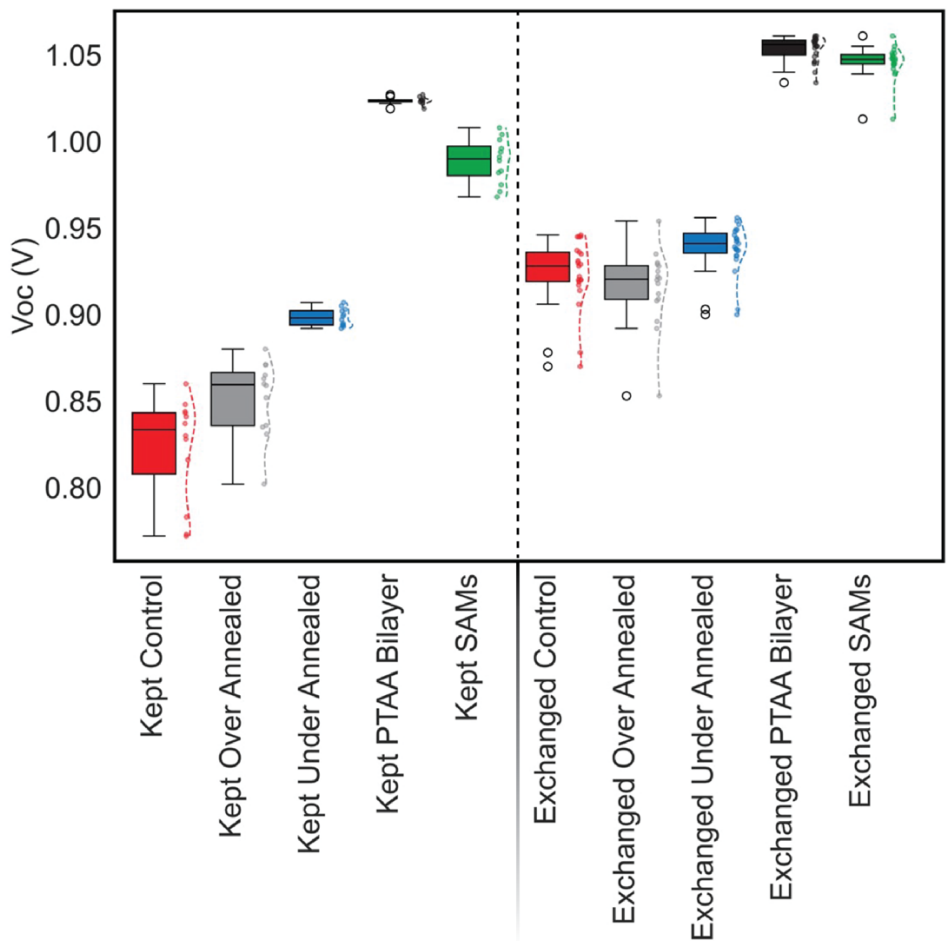

**Figure S2.** Box plots for each batch processing condition showing the device $V_{OC}$ for samples kept at NREL compared to the $V_{OC}$'s of devices exchanged with Optigon. We observe similar trends in the data set with a slight improvement in exchanged device metrics.

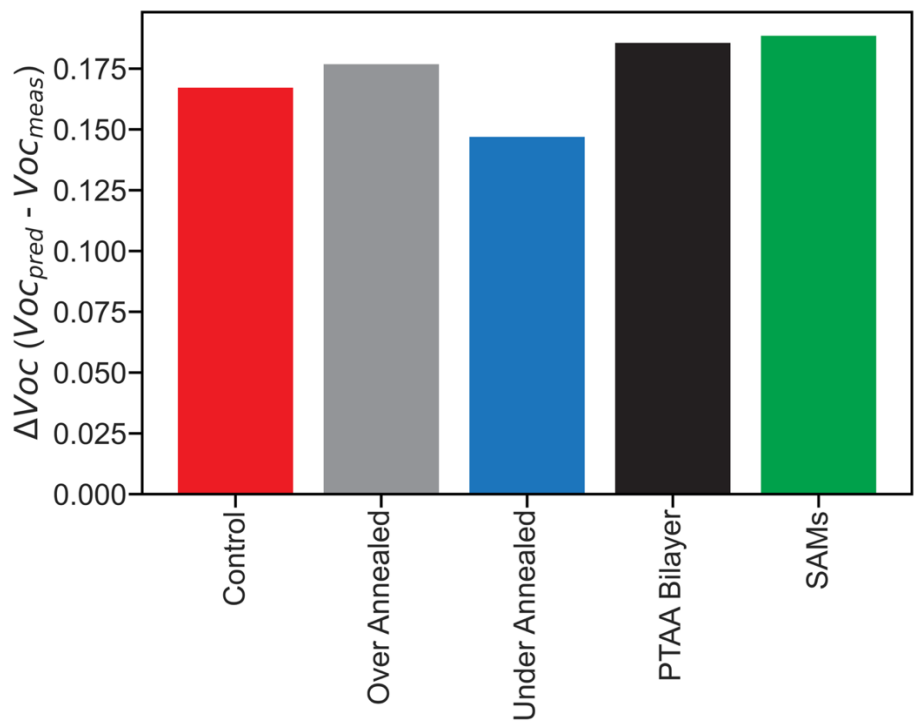

**Figure S3**. Bar graph of the median difference between the predicted $V_{OC}$ ($V_{OC,pred.}$) of the semi-fabricated devices and the measured $V_{OC}$ ($V_{OC,meas.}$) of completed devices. All values are similar, indicating comparable non-radiative loss after introduction of the top charge transport layer and contacts (C60/BCP/Ag), which is expected.